%% file: symbolic.tex
\documentclass[copyright,creativecommons]{eptcs}
\providecommand{\event}{F. Bonchi, D. Grohmann, P. Spoletini, and E. Tuosto:\\ ICE'09 Structured Interactions} 
\providecommand{\volume}{nnn}
\providecommand{\anno}{2009}
\providecommand{\firstpage}{1}
\providecommand{\eid}{3}

\usepackage{breakurl}        
\usepackage{makeidx}  
\usepackage{latexsym,amssymb}
\usepackage{graphicx,color}
\usepackage{times}
\usepackage{theorem}

\newtheorem{theorem}{Theorem}[section]
\newtheorem{definition}[theorem]{Definition}

{
\theoremstyle{plain}
\theorembodyfont{\rmfamily}
\newtheorem{example}[theorem]{Example}
}

\newcommand{\defn}[1]{Def.\,\ref{defn:#1}}
\newcommand{\fig}[2][]{Fig.\,\ref{fig:#2}\ensuremath{#1}}
\newcommand{\tab}[1]{Tab.\,\ref{tab:#1}}
\newcommand{\eq}[1]{(\ref{eq:#1})}

\newcommand{\ex}[1]{Ex.\,\ref{ex:#1}}
\newcommand{\secn}[1]{Sect.\,\ref{sec:#1}}

\newcommand{\mycaption}[1]{\caption{#1}}

\newcounter{tempctr}
\newcommand{\breakenumistart}{%
  \setcounter{tempctr}{\value{enumi}}%
  \end{enumerate}%
}
\newcommand{\breakenumiend}{%
  \begin{enumerate}%
  \setcounter{enumi}{\value{tempctr}}%
}

\font\tensym=msbm10
\font\sevensym=msbm7
\font\fivesym=msbm5

\newfam\symfam
\textfont\symfam=\tensym
\scriptfont\symfam=\sevensym
\scriptscriptfont\symfam=\fivesym
\def\sym{\fam\symfam\tensym}

\newcommand{\sB}{\ensuremath{\mbox{\sym B}}}

\newcommand{\derrule}[3][1]{%
  \ensuremath{%
    \begin{array}{*{#1}{@{\hspace{3mm}}c@{\hspace{3mm}}}}
      #2\\
      \hline 
      \multicolumn{#1}{c}{#3}
    \end{array}%
  }%
}

\newlength{\txtwidth}
\newlength{\temp}

\newcommand{\mdash}[1][]{---{#1}}

\newcommand{\ie}{i.e.\ }
\newcommand{\cf}{cf.\ }

\newcommand{\eg}{e.g.,\ }

\newcommand{\ac}[1][{}]{\mbox{\ensuremath{\cal AC_{#1}}}}

\newcommand{\ct}{\mbox{\ensuremath{\cal CT}}}

\newcommand{\synch}{\cdot}

\newcommand{\goesto}[1][]{\stackrel{#1}{\rightarrow}}
\newcommand{\longgoesto}[1][]{\stackrel{#1}{\longrightarrow}}

\newcommand{\less}{\prec}
\newcommand{\bydef}[1]{\ensuremath{\stackrel{def}{#1}}}
\newcommand{\predactive}{\ensuremath{Act}}

\newcommand{\non}[1]{\ensuremath{\,\overline{#1}}}

\newlength{\lnframe}
\newlength{\lnframep}

\newcommand{\trigger}{
  \put(3,0){\makebox(0,0)[b]{$\blacktriangle$}}
  \put(3,0){\line(0,1){6}}
}

\newcommand{\triggerd}{
  \put(3,0){\makebox(0,0)[t]{$\blacktriangledown$}}
  \put(3,0){\line(0,-1){6}}
}

\newcommand{\synchron}{
  \put(3,3){\circle*{6}}
}

\newcommand{\anytype}{
  \put(3,0){\line(0,1){6}}
}

\newcommand{\ignore}[1]{}

\title{Symbolic Implementation of Connectors in BIP}
\def\titlerunning{Symbolic Implementation of Connectors in BIP}  
\author{Mohamad Jaber \qquad\qquad Ananda Basu
\institute{VERIMAG, Centre {\'E}quation, 
  2 av de Vignate, 38610, Gi{\`er}es, France}
\email{\{Mohamad.Jaber,Ananda.Basu\}@imag.fr}
\and Simon Bluidze
\institute{CEA, LIST, Bo\^{i}te Courrier 94, Gif-sur-Yvette, F-91191 France}
\email{Simon.Bliudze@cea.fr}
}
\def\authorrunning{M. Jaber, A. Basu, and S. Bluidze}   

\begin{document}
\maketitle              

\begin{abstract}        
  BIP is a component framework for constructing systems by superposing
  three layers of modeling: Behavior, Interaction, and Priority.  Behavior
  is represented by labeled transition systems communicating through ports.
  Interactions are sets of ports.  A synchronization between components is
  possible through the interactions specified by a set of connectors.  When
  several interactions are possible, priorities allow to restrict the
  non-determinism by choosing an interaction, which is maximal according to some
  given strict partial order.

  The BIP component framework has been implemented in a language and a
  tool-set.  The execution of a BIP program is driven by a dedicated
  engine, which has access to the set of connectors and priority
  model of the program.  A key performance issue is the computation of the
  set of possible interactions of the BIP program from a given state.

  Currently, the choice of the interaction to be executed involves a costly
  exploration of enumerative representations for connectors.  This leads to
  a considerable overhead in execution times.  In this paper, we propose a
  symbolic implementation of the execution model of BIP, which drastically
  reduces this overhead.  The symbolic implementation is based on computing
  boolean representation for components, connectors, and priorities with an
  existing BDD package.
\end{abstract}

\section{Introduction} \label{sec:intro}

\linespread{0.95}\selectfont

Component-based design is based on the separation between coordination and
computation.  Systems are built from units processing sequential code
insulated from concurrent execution issues.  The isolation of coordination
mechanisms allows a global treatment and analysis.

One of the main limitations of the current state-of-the-art is the lack of
a unified paradigm for describing and analyzing information flow between
components.  Such a paradigm would allow system designers and implementers
to formulate their solutions in terms of tangible, well-founded and
organized concepts instead of using dispersed coordination mechanisms such
as semaphores, monitors, message passing, remote call, protocols etc.  A
unified paradigm should allow a comparison of otherwise unrelated
architectural solutions and could be a basis for evaluating them and
deriving implementations in terms of specific coordination mechanisms.
Furthermore, it should be expressive enough to directly encompass all types
of coordination as discussed in \cite{BliSif08-express-concur}.

A number of paradigms for unifying interaction in heterogeneous systems
have been proposed in \cite{Arbab05,Metro,Sztip,Ptolemy}.  In these works
unification is achieved by reduction to a common low-level semantic model.
Interaction mechanisms and their properties are not studied independently
of behavior.

BIP \cite{bip06} is a component framework for constructing systems by
superposing three layers of modeling: Behavior, Interaction, and Priority.
The lower layer consists of a set of atomic components represented by
transition systems.  The second layer models Interaction between
components.  Interactions are sets of ports specified by connectors
\cite{BliSif08-acp-tc}.  Priority, given by a strict partial order on
interactions, is used to enforce scheduling policies applied to
interactions of the second layer.  The BIP component framework has a formal
operational semantics given in terms of Labeled Transition Systems and
Structural Operational Semantics derivation rules.

The BIP language offers primitives and constructs for modeling and
composing complex behavior from atomic components.  Atomic components are
communicating automata extended with C functions and data.  Transitions are
labeled with sets of communication ports.  Compound components are obtained
from subcomponents by specifying connectors and priorities.

\subsection{Overview of the tool-set} \label{subssec:toolset}

The BIP tool-set\footnote{\
  \texttt{http://www-verimag.imag.fr/\~{}async/bip.php}
} developed at Verimag includes 1)~an editor, for textual
description of systems in the BIP language; 2)~a parser and a deparser, for
generating from a BIP program a model conforming to the BIP
meta-model\footnote{\
  The meta-model is developed in the Eclipse Modeling Framework\\
  (\texttt{http://www.eclipse.org/modeling/emf}).
} and vice versa; and 3)~a compiler for generating executable C++ code.

The execution of a BIP program is driven by a dedicated engine, which has
access to the set of connectors and priority model of the program.
In a given global state, each atomic component waits for an interaction
through a set of active ports (\ie ports labeling active transitions)
communicated to the engine.  The engine computes from the connectors of the
BIP program and the set of all active ports, the set of maximal
interactions (involving active ports).  It chooses one of them, computes
the associated data transfer and notifies the components involved in the chosen
interaction.

The BIP framework has been successfully used for modeling and analysis of a
variety of case studies and applications, such as: performance evaluation
\cite{bip06}, modeling and analysis of TinyOS based wireless sensor network
applications \cite{BMPPS07}, construction and verification of a robotic
system~\cite{RobotBIP}.  In the latter, the code generated by the tool-chain
along with the BIP engine is used as a controller for the robot.  The BIP
model also offers validation techniques for checking essential safety
properties.

\subsection{Problem Definition} \label{subsec:probdef}

A key performance issue is the computation of the set of possible
interactions of the BIP program from a given state.  Currently, the
computation of the maximal interaction involves a costly exploration of
enumerative representations for connectors.  This leads to an important
overhead in execution times.

In this paper, we propose a symbolic implementation of BIP, drastically
reducing this overhead.  This symbolic implementation is based on computing
a boolean representation for components and connectors using an existing
BDD package.\footnote{
  CUDD: CU decision diagram package
  (\texttt{http://vlsi.colorado.edu/\~{}fabio/CUDD/})
}

Sets of interactions in a system with the set of ports $P$ can be
bijectively mapped to the free boolean algebra $\sB[P]$.  Ports are
considered to be boolean variables (\eg for $P=\{p,q\}$, the correspondence
table is shown in \tab{correspondence}).  Whenever an interaction has to be
chosen, a value $true$ or $false$ is assigned to each port variable
according to whether it participates in the interaction or not.  \emph{An
  interaction is possible iff the resulting valuation satisfies the boolean
  function, which describes the interaction model.}

\begin{table}
  \mycaption{Correspondence between sets of interactions and boolean 
    functions with $P = \{p,q\}$}
  \label{tab:correspondence}
  \centering
  \begin{tabular}{c|c}
    \hline\hline
    Sets of interactions $\Big(2^{2^P} \Big)$ & Boolean functions $\Big(\sB[P]\Big)$\\
    \hline
    \renewcommand{\arraystretch}{1.2}
    \begin{tabular}{c}
      $\emptyset$\\
      $\{\emptyset\}\quad\quad \{p\}\quad\quad \{q\}\quad\quad \{p\,q\}$\\
      $\{p,\ \emptyset\}\quad \{q,\ \emptyset\}\quad \{p\,q,\ \emptyset\}\quad \{p,\ q\}\quad \{p,\ p\,q\}\quad \{q,\ p\,q\}$\\
      $\{p,\ q,\ \emptyset\}\quad \{p\,q,\ p,\ \emptyset\}\quad \{p\,q,\ q,\ \emptyset\}\quad \{p\,q,\ p,\ q\}$\\
      $\{p\,q,\ p,\ q,\ \emptyset\}$
    \end{tabular}
    &
    \renewcommand{\arraystretch}{1.2}
    \begin{tabular}{c}
      $false$\\
      
      $\overline{p}\,\overline{q}\quad\quad p\,\overline{q}\quad\quad 
      \overline{p}\,q\quad\quad p\,q$\\
      
      $\overline{q}\quad \overline{p}\quad 
      \overline{p}\,\overline{q} \lor p\,q\quad 
      p\,\overline{q} \lor \overline{p}\,q\quad p\quad q$\\
      
      $\overline{p}\lor \overline{q}\quad p\lor \overline{q}\quad 
      \overline{p}\lor q\quad p \lor q$\\
      
      $true$
    \end{tabular}
    \\\hline\hline
  \end{tabular}
\end{table}

The boolean representation of atomic components and priorities is
relatively simple, whereas a straightforward approach to computing it for
connectors involves costly enumeration of its interactions.  In
\secn{trees}, we present a transformation avoiding this complex enumeration
through a translation from the Algebra of Connectors, $\ac(P)$
\cite{BliSif08-acp-tc}, into the Algebra of Causal Trees, $\ct(P)$
\cite{BliSif08-causal-fmco}.  The terms of the latter have a natural boolean
representation as sets of causal rules (implications).

The paper is structured as follows.  \secn{formal} provides a succinct
presentation of the basic semantic model for BIP and the Algebra of
Connectors. \secn{implem} introduces the boolean representation of atomic
components, connectors, and priorities, which is used for an efficient
implementation of the BIP engine.  \secn{bench} provides the performance
comparison between the current and symbolic implementations.


\section{Formal Semantic Framework} \label{sec:formal}


\subsection{Operational Semantics} \label{sec:sem}

A detailed and fully formalized operational semantics~\cite{BasuBBS08} is beyond the scope
of this paper.  Here, we provide a succinct formalization of the BIP
component model focusing on the operational semantics of component
interaction and priorities, and omitting the aspects related to data
transfer and conditional operation (\ie guards).

\begin{definition}
  Let $P$ be a set of ports.  An {\em interaction} is a subset $a
  \subseteq P$.
\end{definition}


\begin{definition}[Behavior]
  A \emph{labeled transition system} is a triple $B=(Q,P,\goesto)$, where
  $Q$ is a set of {\em states}, $P$ is a set of {\em ports}, and $\goesto\,
  \subseteq Q\times 2^P\times Q$ is a set of {\em transitions}, each
  labeled by an interaction.  We abbreviate $(q,a,q')\in\,\goesto$ to $q
  \goesto[a] q'$.

  An interaction $a$ is {\em active} in a state $q$, denoted $q
  \goesto[a]$, iff there exists $q'\in Q$ such that $q \goesto[a] q'$.  A
  port is active in a state $q$, iff it belongs to an interaction active in
  this state.
\end{definition}

Components in BIP can be \emph{atomic} and \emph{compound}.  Atomic
components are defined by their behavior.  Compound components are obtained
by composition of their subcomponents (atomic or compound) using
interaction and priority models as defined below.

We require that sets of ports of atomic components are pairwise disjoint.
That is, for a system obtained as the composition of $n$ atomic components,
modeled by transition systems $B_i=(Q_i, P_i, \goesto_i)$, for $i\in
[1,n]$, we have $P_i \cap P_j = \emptyset$, for $i,j \in [1,n]$ ($i\not=
j$).

\begin{definition}[Interaction]
  \label{defn:interaction}
  Let $B_i=(Q_i, P_i, \goesto_i)$, for $i\in [1,n]$, be a set of
  components, and $P = \bigcup_{i=1}^n P_i$.  An \emph{interaction model}
  is a set of interactions $\gamma \subseteq 2^P$.

  The component $\gamma(B_1,\dots,B_n)$ is defined by the behavior $(Q, P,
  \goesto_\gamma)$, where $Q = \prod_{i=1}^n Q_i$ and $\goesto_\gamma$ is
  the least set of transitions satisfying the rule
  \begin{equation}
    \label{eq:transsem}
    \derrule[2]{a \in \gamma &
      \forall i\in [1,n],\ (a \cap P_i \not= \emptyset \Rightarrow
      q_i \longgoesto[a \cap P_i]_i q'_i)
    }{
      (q_1,\dots,q_n) \goesto[a]_\gamma (\widetilde{q_1},\dots,\widetilde{q_n})
    }\,,
  \end{equation}
  with $\widetilde{q_i}$ denoting $q_i'$, if $a \cap P_i \neq \emptyset$,
  and $q_i$ otherwise.
\end{definition}

The states of components that do not participate in the interaction remain
unchanged.

\begin{example}[Modulo-8 counter] 
  \label{ex:bipcounter} 
  A BIP model for the modulo-8 counter is shown \fig{bipcounter}.  It has
  three atomic modulo-2 counter components $B_1$, $B_2$, and $B_3$, each
  having an input port (resp. $p$, $r$, and $t$) and an output port (resp.
  $q$, $s$, and $u$).  In a modulo-2 counter, the output port is activated
  on every second activation of the input port.

  The modulo-8 counter is composed by synchronizing the output of $B_1$
  with the input of $B_2$ and the output of $B_2$ with the input of $B_3$.
  This is achieved by the interaction model shown in the figure.
%
\end{example}

\begin{figure}
  \centering
  \begin{picture}(175,80)(-5,-5)
    \put(-5,-5){\framebox(175,80){}}
    \multiput(0,0)(60,0){3}{
      \put(0,0){\framebox(45,45){}}
      \put(2.5,0){
        \put(20,10){\circle{10}}
        \put(20,35){\circle{10}}
        
        \qbezier(16.5,13.5)(10,20)(16.5,31.5)
        \put(14.5,27){\vector(1,2){2.5}}
        
        \qbezier(23.5,13.5)(30,20)(23.5,31.5)
        \put(25.5,18){\vector(-1,-2){2.5}}
      }
    }
    \put(1,1){\makebox(0,0)[bl]{$B_1$}}
    \put(61,1){\makebox(0,0)[bl]{$B_2$}}
    \put(121,1){\makebox(0,0)[bl]{$B_3$}}

    \put(3,0){
    \put(20,10){\makebox(0,0){$\scriptstyle l_{\scriptscriptstyle 1}$}}
    \put(20,35){\makebox(0,0){$\scriptstyle l_{\scriptscriptstyle 2}$}}
    \put(80,10){\makebox(0,0){$\scriptstyle l_{\scriptscriptstyle 3}$}}
    \put(80,35){\makebox(0,0){$\scriptstyle l_{\scriptscriptstyle 4}$}}
    \put(140,10){\makebox(0,0){$\scriptstyle l_{\scriptscriptstyle 5}$}}
    \put(140,35){\makebox(0,0){$\scriptstyle l_{\scriptscriptstyle 6}$}}

    \put(8,20){\makebox(0,0){$p$}}
    \put(32,20){\makebox(0,0){$pq$}}
    \put(68,20){\makebox(0,0){$r$}}
    \put(92,20){\makebox(0,0){$rs$}}
    \put(128,20){\makebox(0,0){$t$}}
    \put(152,20){\makebox(0,0){$tu$}}
  }
    \put(0,30){\framebox(7,10){$p$}}
    \put(38,30){\framebox(7,10){$q$}}
    \put(60,30){\framebox(7,10){$r$}}
    \put(98,30){\framebox(7,10){$s$}}
    \put(120,30){\framebox(7,10){$t$}}
    \put(158,30){\framebox(7,10){$u$}}

    \put(0,50){\framebox(165,20){Interactions: $\{p,\ p q r,\ p q r s t,\ 
	p q r s t u\}$}}
  \end{picture}
  \mycaption{Modulo-8 counter}
  \label{fig:bipcounter}
\end{figure}

An interaction $a \in 2^P$ is active in $\gamma(B_1,\dots,B_n)$ iff, for
all $i \in [1,n]$, $a \cap P_i$ is active in $B_i$ or empty.  An active
interaction $a$ is \emph{enabled} in $\gamma(B_1,\dots,B_n)$ iff $a \in
\gamma$.  For $B=(Q, P, \goesto)$, $q \in Q$, and $a\in 2^P$, we define the
predicate
\begin{equation}
  \label{eq:active}
  \predactive(B,q,a)\  \bydef{=}
  \left\{
    \renewcommand{\arraystretch}{1.3}
    \begin{array}{ll}
      q \goesto[a],
      & \mbox{if $B$ is an atomic behavior},\\
      \lefteqn{\forall\, i \in [1,n], \Big(a \cap P_i\ \neq \emptyset 
        \Rightarrow \predactive(B_i, q_i, a \cap P_i)\Big),}\\
      & \mbox{if $B = \gamma(B_1,\dots,B_n)$ and $q=(q_1,\dots,q_n)$},\\
      \end{array}
    \right.
\end{equation}

Several distinct interactions can be enabled at the same time, thus
introducing non-determinism in the product behavior, which can be
restricted by means of priorities.

\begin{definition}[Priority]
  \label{defn:priority}
  Let $B = (Q,P,\goesto)$ be a behavior.  A {\em priority model} $\pi$ is a
  strict partial order on $2^P$.  Given a priority model $\pi$, we
  abbreviate $(a,a')\in \pi$ to $a \less a'$.

  The component $\pi(B)$ is defined by the behavior $(Q, P, \goesto_\pi)$,
  where $\goesto_\pi$ is the least set of transitions satisfying the rule
  \begin{equation}
    \label{eq:prisem}
    \derrule[2]{
      q \goesto[a] q' &
      \not\exists\,a': \Big(a \less a' \ \land \ \predactive(B,q,a')\Big)
    }{
      q \goesto[a]_\pi q'
    }\,.
  \end{equation}
\end{definition}

An interaction is enabled in $\pi(B)$ only if it is enabled in $B$ and
maximal according to $\pi$ among the active interactions in $B$.

\begin{example}[Sender/Receivers] 
  \label{ex:basicmodels}
  \fig{triv} shows a component $\pi\,\gamma(S,R_1,R_2,R_3)$ obtained by
  composition of four atomic components: a sender $S$, and three
  receivers, $R_1$, $R_2$ and $R_3$. The sender has a port $s$ for sending
  messages, and each receiver has a port $r_i$ ($i=1,2,3$) for receiving
  them.  The second column in \tab{models} specifies $\gamma$ for four
  different coordination schemes listed below.
  \begin{figure}
    \centering
    \begin{picture}(230,95)(-5,-5)
      \put(-5,-5){\framebox(230,95){}}
      \multiput(0,0)(60,0){4}{
        \put(0,0){\framebox(40,45){}}
        \put(15,7){\circle{10}}
        \put(15,37){\circle{10}}
        \put(15,32){\vector(0,-1){20}}
        
        \put(25,35){\framebox(10,10){}}
      }
      
      \put(5,17){\makebox(10,10){$s$}}
      \put(25,35){\makebox(10,10){$s$}}
    
      \put(65,17){\makebox(10,10){$r_1$}}
      \put(85,35){\makebox(10,10){$r_1$}}
      
      \put(125,17){\makebox(10,10){$r_2$}}
      \put(145,35){\makebox(10,10){$r_2$}}
      
      \put(185,17){\makebox(10,10){$r_3$}}
      \put(205,35){\makebox(10,10){$r_3$}}
      
      \put(0,50){\framebox(220,15){Interactions: $\gamma$}}
      \put(0,70){\framebox(220,15){Priorities: $\pi$}}
    \end{picture}
    \caption{A system with four atomic components}
    \label{fig:triv}
  \end{figure}
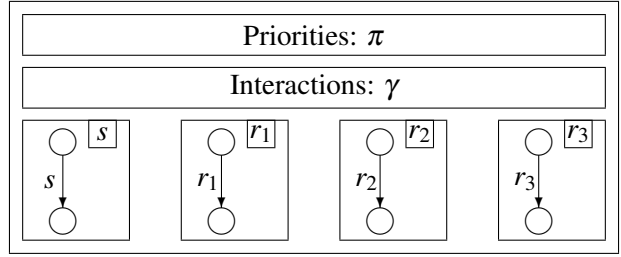
 
  \begin{itemize}
  \item Rendezvous
    means strong synchronization between $S$ and all $R_i$.  This is
    specified by a single interaction involving all the ports.  This
    interaction can occur only if all the components are in states
    enabling transitions labeled respectively by $s$, $r_1$, $r_2$,
    $r_3$.
    
  \item Broadcast
    means weak synchronization, that is a synchronization involving $S$
    and any (possibly empty) subset of $R_i$.  This is specified by the
    set of all interactions containing $s$.  These interactions can
    occur only if $S$ is in a state enabling $s$.  Each $R_i$
    participates in the interaction only if it is in a state enabling
    $r_i$.

  \item Atomic broadcast
    means that either a message is received by all $R_i$, or by none.
    Two interactions are possible: $s$, when at least one of the
    receiving ports is not active, and the interaction
    $s\,r_1\,r_2\,r_3$, corresponding to strong synchronization.

  \item Causal chain
    means that for a message to be received by $R_i$ it has to be
    received at the same time by all $R_j$, for $j<i$.
  \end{itemize}
  
  For rendezvous, the priority model is empty. For all other
  coordination schemes, the \emph{maximal progress} priority model
  ensures that, whenever several interactions are possible, the
  interaction involving a maximal number of ports has higher priority.
  This is achieved by taking $\pi = \{(a,a')\,|\, a\varsubsetneq
  a'\}$.
%
\end{example}

\begin{table}
  \mycaption{Interaction models, connectors and causality trees for 
    four basic coordination schemes} 
  \label{tab:models}

  \centering
  \begin{tabular}[b]{l|l|l|l}
    \hline\hline
    Scheme & Interactions & Connector & Causal Tree
    \\
    \hline
    Rendezvous & 
    $\{s r^{}_1 r^{}_2 r^{}_3\}$ &
    $s r^{}_1 r^{}_2 r^{}_3$  & 
    $s r^{}_1 r^{}_2 r^{}_3$
    \\
    Broadcast & 
    $\{s,\ s r^{}_1,\ s r^{}_2,\ s r^{}_3,\ s r^{}_1 r^{}_2$, &
    $s' r^{}_1 r^{}_2 r^{}_3$ & 
    $s \rightarrow (r^{}_1 \oplus r^{}_2 \oplus r^{}_3)$ \\
    & \quad $s r^{}_1 r^{}_3,\ s r^{}_2 r^{}_3,\ s r^{}_1 r^{}_2 r^{}_3\}$ &
    \\
    Atomic Broadcast & 
    $\{s,\ s r^{}_1 r^{}_2 r^{}_3\}$ &
    $s' [r^{}_1 r^{}_2 r^{}_3]$ &  
    $s \rightarrow r^{}_1 r^{}_2 r^{}_3$
    \\
    Causal Chain & 
    $\{s,\ s r^{}_1,\ s r^{}_1 r^{}_2,\ s r^{}_1 r^{}_2 r^{}_3\}$ &
    $s' [r_1' [r_2' r^{}_3]]$ & 
    $s \rightarrow r_1 \rightarrow r_2 \rightarrow r^{}_3$
    \\
    \hline\hline
  \end{tabular}
\end{table}

\subsection{The Engine} \label{sec:enumprotocol}
The operational semantics is implemented by an engine. In the basic implementation, 
the engine computes the enabled interactions by enumerating over the complete 
list of interactions in the model.

During execution, on each iteration of the engine, the enabled interactions 
are selected from the complete list of interactions, based on the current state 
of the atomic components. Then, between the enabled interactions, priority 
rules are applied to eliminate the ones with low priority.

The main loop of the engine consists of the following steps:
\begin{enumerate}
\item Each atomic component sends to the engine its current state.
\item The engine enumerates on the list of interactions in the model, selects
      the enabled ones based on the current state of the atomic components and
      eliminates the ones with low priority.
\item Amongst the enabled interactions, the engine selects any one and notifies 
      the involved atoms the transition to take.
\end{enumerate}
The time to compute the enabled interactions by engine is proportional to
the number of interactions in the model.


\subsection{The Algebra of Connectors} \label{sec:ac}

In \cite{BliSif08-acp-tc}, the Algebra of Connectors, $\ac(P)$, is
introduced formalizing the concept of connector supported by the BIP
language.  Connectors are used to define the Interactions layer of the
composed system.  They can express complex coordination schemes combining
synchronization by rendezvous and broadcast.

In \cite{BliSif08-acp-tc}, the Algebra of Connectors has two binary
operations: union and fusion.  Union operation allows to combine several
connectors into a single expression, whereas fusion is used to merge two
connectors.  Although here, for the sake of simplicity, we only consider
the $\ac(P)$ terms that do not involve union (\emph{monomial connectors} in
\cite{BliSif08-acp-tc}), the presented results can be extended to the
general case (see \secn{bool:connectors}).

Let $P$ be a set of ports, such that $0,1 \not\in P$. The syntax of the
Algebra of Connectors, $\ac(P)$, is defined by
\begin{equation} \label{eq:acsynt}
  \renewcommand{\arraystretch}{1.3}
  \begin{array}{rcl}
    s & ::= & [0]\ |\ [1]\ |\ [p]\ |\ [x]\mbox{\quad (synchrons)}\\
    t & ::= & [0]'\ |\ [1]'\ |\ [p]'\ |\ [x]'\mbox{\quad (triggers)}\\
    x & ::= & s\ |\ t\ |\ x\synch x\,,
  \end{array}
\end{equation}
for $p\in P$, and where $\synch$ is a binary \emph{fusion} operator, and
brackets $[\cdot]$ and $[\cdot]'$ are unary \emph{typing} operators.

Typing is used to form hierarchically structured connectors: $[\cdot]'$
defines \emph{triggers} (which can initiate an interaction), and $[\cdot]$
defines \emph{synchrons} (which need synchronization with other ports).  In
order to simplify notation, we omit brackets on 0, 1, and ports $p \in P$,
as well as `$\synch$' for the fusion operator.

Complete and formal presentation of the Algebra of Connectors and its
properties can be found in \cite{BliSif08-acp-tc}.  Here we only give the
intuitive semantics.

The semantics of $\ac(P)$ is given in terms of sets of interactions.
Intuitively, it consists in recursively applying the following rule:
\emph{For a connector of the form}
\[[x_1]'\dots[x_n]'[y_1]\dots[y_m]\]
\emph{a possible interaction is a union of interactions from the
  sub-connectors $x_1,\dots,x_n,y_1,\dots,y_m$, comprising an
  interaction from \emph{at least one} of the triggers
  $x_1,\dots,x_n$.  When there are no triggers, an interaction from
  every synchron sub-connector $y_1,\dots,y_m$ is required to form an
  interaction of the complete connector.}

Graphically, connectors are represented as trees with ports at their leaves.
We use triangles and disks to represent types: triggers and synchrons,
respectively.

\begin{example}[Sender/Receivers continued] 
  \label{ex:acbasicmodels}
  The $\ac(P)$ connectors for the four coordination schemes of
  \ex{basicmodels} are given in the third column of \tab{models} and
  illustrated in \fig{connacp}.
\end{example}

\begin{figure}
  \centering
  \begin{tabular}{cccc}
    \begin{picture}(76,25)(-5,0)

      \multiput(0,0)(20,0){4}{
	\put(0,10){\synchron}
	\put(3,16){\line(0,1){9}}
      }
      \put(3,25){\line(1,0){60}}

      \put(3,0){\makebox(0,10){$s$}}
      \put(23,0){\makebox(0,10){$r_1$}}
      \put(43,0){\makebox(0,10){$r_2$}}
      \put(63,0){\makebox(0,10){$r_3$}}
    \end{picture}
    &
    \begin{picture}(76,25)(-5,0)

      \put(0,10){\trigger}
      \put(3,16){\line(0,1){9}}
      \multiput(23,0)(20,0){3}{
	\put(-3,10){\synchron}
	\put(0,16){\line(0,1){9}}
	\put(-20,25){\line(1,0){20}}
      }
     
      \put(3,0){\makebox(0,10){$s$}}
      \put(23,0){\makebox(0,10){$r_1$}}
      \put(43,0){\makebox(0,10){$r_2$}}
      \put(63,0){\makebox(0,10){$r_3$}}
    \end{picture}
    &
    \begin{picture}(71,50)(-5,0)

      \put(0,15){\trigger}
      \put(3,21){\line(0,1){14}}
      \multiput(18,0)(20,0){3}{
	\put(-3,10){\synchron}
	\put(0,16){\line(0,1){9}}
      }
      \put(18,25){\line(1,0){40}}

      \put(35,25){\synchron}
      \put(38,31){\line(0,1){4}}
      \put(3,35){\line(1,0){35}}

      \put(3,5){\makebox(0,10){$s$}}
      \put(18,0){\makebox(0,10){$r_1$}}
      \put(38,0){\makebox(0,10){$r_2$}}
      \put(58,0){\makebox(0,10){$r_3$}}
    \end{picture}
    &
    \begin{picture}(83,50)(-7,0)

      \put(38,10){
	\put(-3,0){\trigger}
	\put(27,0){\synchron}
	\put(0,6){\line(0,1){9}}
	\put(30,6){\line(0,1){9}}
	\put(0,15){\line(1,0){30}}
      }
      \multiput(13,25)(-17,10){2}{
	\put(2,-10){\trigger}
	\put(37,0){\synchron}
	\put(40,6){\line(0,1){4}}
	\put(5,-4){\line(0,1){14}}
	\put(5,10){\line(1,0){35}}
      }
      
      \put(3,15){\makebox(0,10){$s$}}
      \put(18,5){\makebox(0,10){$r_1$}}
      \put(38,0){\makebox(0,10){$r_2$}}
      \put(68,0){\makebox(0,10){$s_3$}}
    \end{picture}
    \\
    \emph{Rendezvous} & \emph{Broadcast} &
    \emph{Atomic broadcast} & \emph{Causal chain}
  \end{tabular}
  \mycaption{Graphic representation of connectors}
  \label{fig:connacp}
\end{figure}


\subsection{Causal Trees} \label{sec:trees}

In \cite{BliSif08-causal-fmco}, the Algebra of Causal Trees, $\ct(P)$, is
introduced, which allows the efficient computation of the boolean
representation of the $\ac(P)$ connectors.  This is achieved by rendering
explicit the causal dependencies between ports participating in the
interactions of a given $\ac(P)$ connector.

In a fusion of typed connectors, triggers are mutually independent, and can
be considered \emph{parallel} to each other.  Synchrons participate in an
interaction only if it is initiated by a trigger.  This introduces a causal
relation: the trigger is a \emph{cause} that can provoke an \emph{effect},
which is the participation of a synchron in an interaction.  For example,
for connectors involving ports $p$ and $q$, there are essentially three
possibilities:
\begin{enumerate}
\item A strong synchronization $pq$.
\item One trigger $p'q$, \ie $p$ is the cause of an interaction and $q$ a
  potential effect, which we will denote in the following by $p \rightarrow
  q$.
\item Two triggers $p'q'$, \ie $p$ and $q$ are independent (parallel),
  which we will denote in the following by $p \oplus q$.
\end{enumerate}

The syntax of the {\em Algebra of Causal Trees}, $\ct(P)$, is defined by
\begin{equation}
  \label{eq:ctsyn}
  t ::= a \,|\, t \rightarrow t \,|\, t\oplus t\,,
\end{equation}
where $a \in 2^P$ is an interaction, $\rightarrow$ and $\oplus$ are
respectively the {\em causality} and the {\em parallel composition}
operators.  Causality binds stronger than parallel composition.
  
Although the causality operator is not associative, for $t_1,\dots,t_n \in
\ct(P)$, we abbreviate $t_1 \rightarrow (t_2 \rightarrow (\dots \rightarrow
t_n)\dots))$ to $t_1 \rightarrow t_2 \rightarrow \dots \rightarrow t_n$.
We call this construction a {\em causal chain}.

Complete and formal presentation of the Algebra of Causal Trees and its
properties can be found in \cite{BliSif08-causal-fmco}.  Here we only give
the intuitive semantics.

The semantics of $\ct(P)$ is given in terms of sets of interactions.
Intuitively, it consists in applying the causal rule: \emph{For a node of a
  causal tree to participate in an interaction, it is necessary that its
  parent participate also.}

\begin{example}[Sender/Receivers continued]
  \label{ex:ctbasicmodels} 
  The causal trees for the four coordination schemes of the
  \ex{basicmodels} are given in the last column of \tab{models} and
  illustrated in \fig{connctp}.
\end{example}

\begin{figure}
  \centering
  \begin{tabular}{c*{3}{@{\hspace{10mm}}c}}
    $s r_1 r_2 r_3$
    &
    \begin{minipage}{52\unitlength}
      \centering
      \begin{picture}(52,35)(-26,0)
      
        \put(0,35){\makebox(0,0)[ct]{$s$}}
        \put(-25,0){\makebox(0,0)[lb]{$r_1$}}
        \put(0,0){\makebox(0,0)[cb]{$r_2$}}
        \put(26,0){\makebox(0,0)[rb]{$r_3$}}

        \put(0,29){\vector(-1,-1){20}}
        \put(0,29){\vector(0,-1){20}}
        \put(0,29){\vector(1,-1){20}}
      \end{picture}
    \end{minipage}
    &
    \begin{minipage}{30\unitlength}
      \centering
      \begin{picture}(30,35)(-15,0)
      
        \put(0,35){\makebox(0,0)[ct]{$s$}}
        \put(0,0){\makebox(0,0)[cb]{$r_1\, r_2\, r_3$}}

        \put(0,28){\vector(0,-1){18}}
      \end{picture}
    \end{minipage}
    &                           
    \begin{minipage}{75\unitlength}
      \begin{picture}(75,45)

        \put(0,45){\makebox(0,0)[lt]{$s$}}
        \put(24,30){\makebox(0,0)[cc]{$r_1$}}
        \put(48,17){\makebox(0,0)[cc]{$r_2$}}
        \put(76,0){\makebox(0,0)[rb]{$r_3$}}

        \put(7,41){\vector(3,-2){10}}
        \put(31,27){\vector(3,-2){10}}
        \put(55,13){\vector(3,-2){10}}
      \end{picture}
    \end{minipage}
    \\
    \emph{Rendezvous} & \emph{Broadcast} &
    \emph{Atomic broadcast} & \emph{Causal chain}
  \end{tabular}
  \mycaption{Causal trees representation of connectors}
  \label{fig:connctp}
\end{figure}

The function $\tau : \ac(P) \rightarrow \ct(P)$ associating a causal tree
with a connector is defined by the following equations:
\begin{eqnarray}
  \tau(p) & = & p\,,\\
  \label{eq:contree1}
  \tau\left([x]'\,\prod_{i=1}^n [y_i]\right) & = & \tau(x) \rightarrow
     \bigoplus_{i=1}^n \tau(y_i)\,,\\
  \label{eq:contree2}
  \tau\Big([x_1]'[x_2]'\Big) & = &
     \tau(x_1) \oplus \tau(x_2)\,,\\
  \label{eq:contree0}
  \tau\Big([y_1][y_2]\Big) & = &
     \bigoplus_{i=1}^{m_1} \bigoplus_{j=1}^{m_2}
     a^1_i a^2_j \rightarrow \Big(t^1_i \oplus t^2_j\Big)\,,
\end{eqnarray}
where $x,x_1,x_2,y_1,\dots,y_n \in \ac(P)$, $p \in P \cup \{0,1\}$,
and, in \eq{contree0}, we assume $\tau(y_k) = \bigoplus_{i=1}^{m_k}
a^k_i \rightarrow t^k_i$, for $k=1,2$.

The equations above are sufficient to define $\tau$.  Indeed, denoting by
`$\simeq$' the equivalence relation induced on $\ac(P)$ by the semantics
presented in \secn{ac}, one can observe that
\begin{eqnarray*}
  [x_1]'\dots[x_n]'[y_1]\dots[y_m]
  & \simeq &
  \left[\Big[\!\dots\!\Big[[x_1]'[x_2]'\Big]'\dots\Big]'[x_n]'\right]'
       [y_1]\dots[y_m]
  \\
  {}[y_1]\dots[y_m]
    & \simeq &
  \left[\Big[\!\dots\!\Big[[y_1][y_2]\Big]\dots\Big][y_{m-1}]\right][y_m]\,,
\end{eqnarray*}
which means that for any connector in $\ac(P)$ there is a uniquely defined
equivalent representation, to which one of the equations
\eq{contree1}--\eq{contree0} is applicable directly.

\begin{example}
  \label{ex:trees}
  Consider $P=\{p,q,r,s,t,u\}$ and $p'q'\Big[[r's] [t' u]\Big] \in \ac(P)$.
  We have
  \begin{eqnarray*}
    \lefteqn{\tau\left(p'q'\Big[[r's] [t' u]\Big]\right)
    \ \bydef{=}\ \tau\left(\Big[p'q'\Big]'\Big[[r's] [t'u]\Big]\right)
    \ =\ \tau(p'q') \rightarrow \tau\Big([r's] [t'u]\Big)}\\
    & = & (p \oplus q) \rightarrow \Big(rt \rightarrow (s \oplus u)\Big)
    \ =\ \Big(p \rightarrow rt \rightarrow (s \oplus u)\Big)
    \ \oplus\ \Big(q \rightarrow rt \rightarrow (s \oplus u)\Big)\,.
  \end{eqnarray*}
\end{example}


\section{Symbolic Implementation} \label{sec:implem}

In the enumerative BIP engine, for each connector, the engine needs to
compute all the possible interactions, check which ones are enabled in the
current global state of the system, and select a maximal enabled one to be
executed.  As interactions are sets of ports, their number is potentially
exponential in the number of ports in the connector. Hence, in the worst
case, the performance of this engine can be extremely poor.

The boolean BIP engine leverages on representing component behavior,
connector interactions, and priorities as boolean functions.\footnote{\
  The implementation of the boolean functions is made using the BDD package
  CUDD.
} For an atomic component, all ports and control states are represented by
boolean variables.  This allows to encode behavior as a boolean expression
of these variables.  Similarly, each connector is represented by the
boolean expression on its ports.  The global behavior is obtained as a
boolean operation on the expressions representing atomic components,
connectors, and priorities.

The choice of an interaction to be executed boils down to evaluating the
control states, substituting their respective boolean variables, and
picking a valuation of the port variables satisfying the boolean expression
that represents the global behavior.

The boolean representation of connectors replaces the costly enumeration
step by efficient BDD manipulations.  In comparison to the exponential cost
of the enumerative engine, this renders a more efficient engine with
evaluation that, in practice, remains linear.  The following sections
describe in detail the boolean representation of the individual BIP
elements and its evaluation by the engine.


\subsection{Atomic Components} \label{sec:bool:atom}

For each atomic component $B_i = (Q_i,P_i,\goesto)$ and each state $q \in
Q_i$, we define boolean functions $f_q, f_{B_i} \in \sB[Q_i, P_i]$
\begin{equation}
  f_q  =  q \land \bigwedge_{q' \neq q} \non{q'}\ \land 
  \bigvee_{q \goesto[a]} \left(a \land 
    \bigwedge_{p \in P_i \setminus a}\non{p}\right)\,,
  \quad
  f_{B_i}  =  \bigvee_{q \in Q_i} f_q \lor \bigwedge_{p \in P_i}\non{p} \,.
\end{equation}

There are two possibilities for a valuation on $Q_i$ and $P_i$ satisfying
$f_{B_i}$: 1)~exactly one state variable $q \in Q_i$ is set to $true$ and
valuations of port variables in $P_i$ correspond to possible transitions of
$B_i$ from the state $q$; 2)~ all port variables are set to $false$, which
means that the component does not change its state.

The boolean function, representing all the possible transitions of the
product automaton, is then the conjunction $f_B = \bigwedge_{i=1}^n
f_{B_i}$.

\begin{example}
  Consider the first Modulo-2 counter component in the Modulo-8 counter
  (\ex{bipcounter}).  To avoid confusion in notations, we denote the states
  of this component by $l_1$ and $l_2$.  The boolean function representing
  this component is then $f_{B_1} = l_1 \non{l_2} p \non{q} \lor \non{l_1}
  l_2 p q \lor \non{p} \non{q}$, and the functions representing the other
  two Modulo-2 counters are computed similarly.  Taking the conjunction, we
  obtain the boolean function representing the product of the three atomic
  components:
 
  \begin{equation}
    \label{eq:atom}
    f_B  = 
    (l_1 \non{l_2}\, p \non{q} \lor  \non{l_1}\, l_2\, p\, q \lor \non{p} \non{q}) \land
    (l_3 \non{l_4}\, r \non{q} \lor  \non{l_3}\, l_4\, r\, s \lor \non{r} \non{s})
    \land\ 
    (l_5 \non{l_6}\, t \non{q} \lor  \non{l_5}\, l_6\, t\, u \lor \non{t} \non{u})\,.    
  \end{equation}
\end{example}


\subsection{Connectors} \label{sec:bool:connectors}

Connectors are represented in boolean form as conjunctions of \emph{causal
  rules} \cite{BliSif08-causal-fmco}. A causal rule is a boolean formula
over a set of ports taking the form of an implication $p \Rightarrow
\bigvee_{i=1}^n a_i$, where $p$ is a port and $a_i$, for $i \in [1,n]$, are
monomials representing interactions.  An interaction $a$ satisfies a causal
rule if the valuation that it defines on the set of ports satisfies the 
boolean expression defining the causal rule. Notice, that this also means
that either $p$ does not belong to $a$ or at least one of $a_i$ does, \ie
$a \models p \Rightarrow \bigvee_{i=1}^n a_i$ iff $p \in a \Rightarrow
\exists i \in [1,n]: a_i \subseteq a$.

Assume now that $P$ is the set of all ports in the system.  In order to
obtain a boolean representation for connectors, we first compute, for each
connector $x$, the corresponding causal tree $t = \tau(x)$ (\cf
\secn{trees}).  The boolean function $f_C \in \sB[P]$ representing a
connector is the conjunction of causal rules obtained from the causal tree
essentially by inverting the arrows and the disjunction of roots of $t$
(see \ex{countercntd} below).  Adding this last disjunction ensures that at
least one of the nodes forming the roots of causal trees participate in the
interaction.  

\begin{example}
  \label{ex:countercntd}
  The connector $x = p'[[q\,r]'[[s\,t]' u]]$ realizes exactly the
  interaction model of the \ex{bipcounter} (\cf
  \cite{BliSif08-acp-tc}).  The corresponding causal tree is $\tau(x) =
  p \rightarrow qr \rightarrow st \rightarrow u$.  Inverting the arrows and
  taking in account strong synchronizations (\eg for $q$ to participate in
  the interaction, $r$ also has to participate), we obtain the causal rules
  \[
  q \Rightarrow pr,\quad
  r \Rightarrow pq,\quad
  s \Rightarrow qrt,\quad
  t \Rightarrow qrs,\quad
  u \Rightarrow st\,.
  \]
  The boolean function representing this connector is the conjunction of
  these rules together with $p$, which is the root of $\tau(x)$:
  \[
    f_x = p \land
    (q \Rightarrow pr) \land
    (r \Rightarrow pq) \land 
    (s \Rightarrow qrt) \land
    (t \Rightarrow qrs) \land
    (u \Rightarrow st)\,.
  \]
\end{example}

For a system having several connectors $C_1,\dots,C_m$, boolean functions
are individually computed as above for each of the $C_i$ and combined by
taking $f_C = \bigvee_{i=1}^m \left(f_{C_i} \land \bigwedge_{p \not\in
    C_i}\non{p}\right)$, where $p \not\in C_i$ means that the port $p$ is
not used in $C_i$.


\subsection{Priorities} \label{sec:bool:priority}

To obtain a boolean representation of a priority model, given by a strict
partial order $\pi$ on $2^P$, we define a function $f_P \in \sB[P,P']$,
where $P' = \{p'|p \in P\}$.\footnote{
  The primes here are not related to those in the $\ac(P)$ syntax.
} Abbreviating $(a,a') \in \pi$ to $a \less a'$, we put
\begin{eqnarray}
  \label{eq:priority}
  f_P & = & \bigvee_{a\less a'} \left(
    \bigwedge_{p \in a} p \land
    \bigwedge_{p \not\in a} \non{p} \land
    \bigwedge_{p \in a'} p' \land
    \bigwedge_{p \not\in a'} \non{p'}
    \right)\,.
\end{eqnarray}
Clearly, a pair of interactions $(a,a')$ belongs to $\pi$ (\ie $a$ has
lower priority than $a'$) iff the corresponding boolean valuation $(a,a')$
satisfies $f_P$.

Notice that, according to \eq{prisem}, in a given state $q$ of a behaviour
$B$, a priority is only applied if the triple $(B,q,a')$ satisfies the
predicate $\predactive$.  However, it can be easily verified that
\begin{eqnarray*}
  \predactive(B,q,a') & \Longleftrightarrow &
  (a', q) \models f_B \land \bigwedge_{i=1}^n q_i\,,
\end{eqnarray*}
where the conjunction in the right-hand side represents the global state of
the system, and $f_B$ is the function computed in \eq{atom}.


\subsection{The Engine Protocol} \label{sec:protocol}

The following protocol is used at each step of the execution to choose an
interaction to be fired.  It starts with an initialization phase, where the
following boolean functions are computed: $f_B \in \sB[\,\bigcup_{i=1}^n
Q_i,P]$ representing the atomic components; $f_C \in \sB[P]$ representing
the connectors; and $f_P \in \sB[P,P']$ representing the priorities.  The
conjunction $f_{S} = f_B \land f_C$ is also computed at this stage.

The main loop of the engine consists of the following steps:
\begin{enumerate}
\item Each atomic component $B_i$ sends to the engine its current state
  $q_i \in Q_i$.
  
\item The engine picks any valuation $a$ on $P$, such that
  \begin{equation}
    \label{eq:protocol} 
    \left((a,q) \models f_S \land \bigwedge_{i=1}^n q_i\right)
    \ \ \land\ \  \not\exists\, a': 
    \Big((a, a', q)\models  f_P \land f_B[x'/x] \land  
    \bigwedge_{i=1}^n q_i\Big)\,,
  \end{equation}
  where $q$ is the valuation on $\bigcup_{i=1}^n Q_i$ representing the
  global state of the system and $a'$ is a valuation on $P'$ (\cf
  \secn{bool:priority}).  The function $f_B[x'/x]$ is obtained by
  substituting $p'$, for each port variable $p$ in $f_B$.

\item The engine notifies components by communicating to each $B_i$ the
  label $a \cap P_i$ of the transition to take.
\end{enumerate}  
  
In \eq{protocol}, $(a,q) \models f_S$ implies $a \models f_C$, which means
that $a \in \gamma$, \ie the interaction $a$ is authorized by the
interaction model (\cf \defn{interaction}).  Similarly, $(a,a',q) \models
f_B$ and $(a, a',q) \models f_B[x'/x]$ mean that $a$ and $a'$ respectively
are active in the current global state $q$ of the system.  Finally, $(a,
a',q)\models f_P$ means $a \less a'$.  Thus, any interaction $a$, satisfying
\eq{protocol}, represents an enabled interaction.


\subsection{Complexity of the Engine Protocol} \label{sec:complexity}

Observe that the BDDs for all the involved functions ($f_B$, $f_C$, $f_S$,
and $f_P$) are only computed once and remain constant throughout the
execution of the BIP model.  Thus, the only operations performed at each
iteration of the engine loop are the conjunctions and the existence check
in \eq{protocol}.

First consider the conjunction of $f_S$ with state variables representing
current states of atomic components.  In general, the complexity of
computing a conjunction of two BDDs is proportional to the product of their
sizes \cite{bryant86}.  Consequently, the complexity of the restriction of
a BDD by substituting a value for one of the variables (\eg computing $f_S
\land q_i$) is linear in the size of the BDD.  However, it can also be
shown that the complexity of taking the conjunction ${f_S} \land
\bigwedge_{i=1}^n q_i$ is also linear in the size of $f_S$.  This is due to
two reasons: 1)~the BDD $\bigwedge_{i=1}^n q_i$ encoding the current state
of the system has one node for each atomic component (see \fig[(b)]{bdds}),
and 2)~the variables $q_1,\dots,q_n$ appear in the same order in the BDDs
for both $f_S$ and $\bigwedge_{i=1}^n q_i$.  Thus, when the size of the BDD
for $f_S$ is small, so is the complexity of the symbolic engine.

\begin{figure}
  \centering
  \begin{tabular}{c@{\hspace{20mm}}c}    
    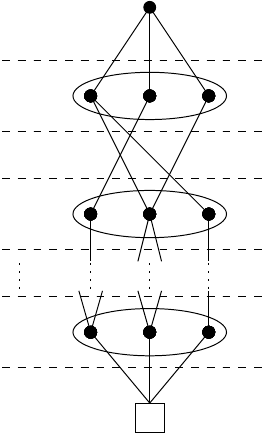
    &
    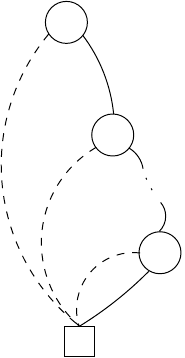
    \\
    $(a)$ & $(b)$
  \end{tabular}
  \caption{Schematic representation of the $f_S$ BDD $(a)$ and the BDD
    $\bigwedge_{i=1}^n q_i$ representing the current state of the system
    $(b)$}
  \label{fig:bdds}
\end{figure}

When constructing the BDD for $f_S$, we alternate the port and control
state variables, as presented schematically in \fig[(a)]{bdds}.  Indeed, it
is well known that the order of variables in a BDD has an important
influence on the complexity of boolean operations.  Clearly, the variables
representing ports and control states of an atomic component are strongly
correlated.

As the main goal of this paper is to demonstrate the feasibility of the
presented approach, the current implementation is limited to the boolean
representation of atomic components and connectors.  It does not include
priorities.  Observe, however, that the combination of the existence check
with the conjunction of BDDs, in \eq{protocol}, has been widely used for
symbolic model checking \cite{clarke92}.  The same argument as above shows
that the complexity of the conjunction is linear in the size of $f_P \land
f_B[x'/x]$ (this BDD can also be precomputed), whereas the complexity of
the existence check is constant.


\section{Experimental Results} 
\label{sec:bench}

We compare the engine execution times of the enumerative and boolean
engines for two examples.  The BIP models for both examples are limited to
synchronization, \ie do not have any data transfer.  We present below the
two examples and the simulation results.


\subsection{Bus} 
\label{sec:bus}

In this example, we consider a system consisting of $N$ independent
clusters of components communicating through a ``bus'', \ie a single common
connector (see \fig{bus}).  Each cluster consists of four components that
alternate some computation (transitions labeled $c_i$, for $i \in [1,4]$)
and communication (transitions labeled $s_i$, for $i \in [1,4]$).

\begin{figure}
  \centering
  \begin{picture}(230,71)(0,-6)
    \multiput(0,0)(60,0){4}{
      \put(0,0){\framebox(50,50){}}
      \put(2.5,5){      
        \put(10,20){\circle{10}}
        \put(35,20){\circle{10}}
        
        \qbezier(13.5,16.5)(20,10)(31.5,16.5)
        \put(22.5,13.75){\vector(1,0){2.5}}
        
        \qbezier(13.5,23.5)(20,30)(31.5,23.5)
        \put(22.5,26.5){\vector(-1,0){2.5}}
      }
      \put(25,56){\line(0,1){9}}
      \put(22,0){\triggerd}
    }

    \put(202,50){\synchron}
    \multiput(0,0)(60,0){3}{      
      \put(22,50){\trigger}
      \put(25,65){\line(1,0){60}}
    }    

    \put(29,34){\makebox(0,0)[l]{$\scriptstyle s^{}_{\scriptscriptstyle 1}$}}
    \put(22,15){\makebox(0,0)[r]{$\scriptstyle c^{}_{\scriptscriptstyle 1}$}}
    \put(89,34){\makebox(0,0)[l]{$\scriptstyle s^{}_{\scriptscriptstyle 2}$}}
    \put(82,15){\makebox(0,0)[r]{$\scriptstyle c^{}_{\scriptscriptstyle 2}$}}
    \put(149,34){\makebox(0,0)[l]{$\scriptstyle s^{}_{\scriptscriptstyle 3}$}}
    \put(142,15){\makebox(0,0)[r]{$\scriptstyle c^{}_{\scriptscriptstyle 3}$}}
    \put(209,34){\makebox(0,0)[l]{$\scriptstyle s^{}_{\scriptscriptstyle 4}$}}
    \put(202,15){\makebox(0,0)[r]{$\scriptstyle c^{}_{\scriptscriptstyle 4}$}}

    \put(20,40){\framebox(10,10){$s^{}_1$}}
    \put(20,0){\framebox(10,10){$c^{}_1$}}
    \put(80,40){\framebox(10,10){$s^{}_2$}}
    \put(80,0){\framebox(10,10){$c^{}_2$}}
    \put(140,40){\framebox(10,10){$s^{}_3$}}
    \put(140,0){\framebox(10,10){$c^{}_3$}}
    \put(200,40){\framebox(10,10){$s^{}_4$}}
    \put(200,0){\framebox(10,10){$c^{}_4$}}
  \end{picture}
  \mycaption{A unit cluster for the Bus example}
  \label{fig:bus}
\end{figure}

Computations of the four components in a cluster are completely independent
and cannot be synchronized.  Thus, for each $i \in [1,4]$, there is a
singleton connector $c_i$.  On the other hand, communications $s_i$ are
weakly synchronized by the connector $s_1' s_2' s_3' s_4$.  In this
connector, the ports $s_1$, $s_2$, and $s_3$ are triggers, whereas $s_4$ is
a synchron.  This means that communication is only possible through $s_4$
when at least one other component is ready to communicate\mdash the fourth
component is an \emph{observer}.

A system with $N$ clusters (\ie $4N$ components) has $5N$ connectors.  We
say that connectors are \emph{sparse} in this system, which favors the
enumerative engine as its execution time becomes linear in the number of
components.


\subsection{Preemptable Tasks} 
\label{sec:task}

This example originates from \cite{WDES}, where a performance evaluation
problem is considered with timed tasks running concurrently on shared
processors.  Here, we disregard the timed aspects of this example and only
consider the task behavior concerned with processor sharing.

Consider $M$ processors and $N$ tasks that can be executed on any
processor.  A processor can have at most two tasks assigned to it at a
time: one running and one preempted.  On arrival of a new task, the running
one is preempted.  A task is resumed, when the one that has preempted it terminates.

\begin{figure}
 \centering
 \begin{tabular}{c*{2}{@{\hspace{1cm}}c}}
   \begin{picture}(90,90)
     \put(0,0){\framebox(90,90){}}
     \put(45,45){\circle{10}}
     \put(45,45){\makebox(0,0){$\scriptstyle s$}}
     \put(45,63){\circle{10}}
     \put(45,63){\makebox(0,0){$\scriptstyle c_1$}}
     \put(45,81){\circle{10}}
     \put(45,81){\makebox(0,0){$\scriptstyle w_1$}}
     \put(63,45){\circle{10}}
     \put(63,45){\makebox(0,0){$\scriptstyle c_4$}}
     \put(81,45){\circle{10}}
     \put(81,45){\makebox(0,0){$\scriptstyle w_4$}}
     \put(27,45){\circle{10}}
     \put(27,45){\makebox(0,0){$\scriptstyle c_2$}}
     \put(9,45){\circle{10}}
     \put(9,45){\makebox(0,0){$\scriptstyle w_2$}}
     \put(45,27){\circle{10}}
     \put(45,27){\makebox(0,0){$\scriptstyle c_3$}}
     \put(45,9){\circle{10}}
     \put(45,9){\makebox(0,0){$\scriptstyle w_3$}}

     \put(55,82){\framebox(8,8){$b_1$}}
     \put(68,82){\framebox(8,8){$f_1$}}
     \put(25,82){\framebox(8,8){$p_1$}}
     \put(12,82){\framebox(8,8){$r_1$}}

     \put(82,55){\framebox(8,8){$b_4$}}
     \put(82,68){\framebox(8,8){$f_4$}}
     \put(82,25){\framebox(8,8){$p_4$}}
     \put(82,12){\framebox(8,8){$r_4$}}
     
     \put(55,0){\framebox(8,8){$b_3$}}
     \put(68,0){\framebox(8,8){$f_3$}}
     \put(25,0){\framebox(8,8){$p_3$}}
     \put(12,0){\framebox(8,8){$r_3$}}
     
     \put(0,55){\framebox(8,8){$b_2$}}
     \put(0,68){\framebox(8,8){$f_2$}}
     \put(0,25){\framebox(8,8){$p_2$}}
     \put(0,12){\framebox(8,8){$r_2$}}

     \qbezier(48,48)(49,50)(50,52)   
     \put(50,52){\vector(0,1){4}}
     \qbezier(50,56)(49,58)(48,60)
     \put(50,57){\makebox(0,0)[l]{$\scriptstyle b^{}_{\scriptscriptstyle 1}$}}
     
     \qbezier(42,48)(41,50)(40,52)   
     \put(40,56){\vector(0,-1){4}}
     \qbezier(40,56)(41,58)(42,60)
     \put(33,58){\makebox(0,0)[l]{$\scriptstyle f^{}_{\scriptscriptstyle 1}$}}
     
     \qbezier(48,66)(49,68)(50,70)   
     \put(50,70){\vector(0,1){4}}
     \qbezier(50,74)(49,76)(48,78)
     \put(52,70){\makebox(0,0)[l]{$\scriptstyle p^{}_{\scriptscriptstyle 1}$}}
     
     \qbezier(42,66)(41,68)(40,70)   
     \put(40,74){\vector(0,-1){4}}
     \qbezier(40,74)(41,76)(42,78)
     \put(32,70){\makebox(0,0)[l]{$\scriptstyle r^{}_{\scriptscriptstyle 1}$}}

     \qbezier(48,42)(49,40)(50,38)   
     \put(50,34){\vector(0,1){4}}
     \qbezier(50,34)(49,32)(48,30)
     \put(50,32){\makebox(0,0)[l]{$\scriptstyle f^{}_{\scriptscriptstyle 3}$}}
     
     \qbezier(42,42)(41,40)(40,38)   
     \put(40,38){\vector(0,-1){4}}
     \qbezier(40,34)(41,32)(42,30)
     \put(33,32){\makebox(0,0)[l]{$\scriptstyle b^{}_{\scriptscriptstyle 3}$}}
     
     \qbezier(48,24)(49,22)(50,20)   
     \put(50,16){\vector(0,1){4}}
     \qbezier(50,16)(49,14)(48,12)
     \put(52,16){\makebox(0,0)[l]{$\scriptstyle r^{}_{\scriptscriptstyle 3}$}}
     
     \qbezier(42,24)(41,22)(40,20)   
     \put(40,20){\vector(0,-1){4}}
     \qbezier(40,16)(41,14)(42,12)
     \put(32,16){\makebox(0,0)[l]{$\scriptstyle p^{}_{\scriptscriptstyle 3}$}}

     \qbezier(48,48)(50,49)(52,50)   
     \put(52,50){\vector(1,0){4}}
     \qbezier(56,50)(58,49)(60,48)
     \put(55,52){\makebox(0,0)[l]{$\scriptstyle b^{}_{\scriptscriptstyle 4}$}}
     
     \qbezier(48,42)(50,41)(52,40)   
     \put(56,40){\vector(-1,0){4}}
     \qbezier(56,40)(58,41)(60,42)
     \put(55,36){\makebox(0,0)[l]{$\scriptstyle f^{}_{\scriptscriptstyle 4}$}}
     
     \qbezier(66,48)(68,49)(70,50)   
     \put(70,50){\vector(1,0){4}}
     \qbezier(74,50)(76,49)(78,48)
     \put(70,54){\makebox(0,0)[l]{$\scriptstyle p^{}_{\scriptscriptstyle 4}$}}
     
     \qbezier(66,42)(68,41)(70,40)   
     \put(74,40){\vector(-1,0){4}}
     \qbezier(74,40)(76,41)(78,42)
     \put(70,36){\makebox(0,0)[l]{$\scriptstyle r^{}_{\scriptscriptstyle 4}$}}
     
     \qbezier(42,48)(40,49)(38,50)   
     \put(34,50){\vector(1,0){4}}
     \qbezier(34,50)(32,49)(30,48)
     \put(28,53){\makebox(0,0)[l]{$\scriptstyle f^{}_{\scriptscriptstyle 2}$}}
     
     \qbezier(42,42)(40,41)(38,40)   
     \put(38,40){\vector(-1,0){4}}
     \qbezier(34,40)(32,41)(30,42)
     \put(28,37){\makebox(0,0)[l]{$\scriptstyle b^{}_{\scriptscriptstyle 2}$}}
     
     \qbezier(24,48)(22,49)(20,50)   
     \put(16,50){\vector(1,0){4}}
     \qbezier(16,50)(14,49)(12,48)
     \put(12,54){\makebox(0,0)[l]{$\scriptstyle r^{}_{\scriptscriptstyle 2}$}}
     
     \qbezier(24,42)(22,41)(20,40)   
     \put(20,40){\vector(-1,0){4}}
     \qbezier(16,40)(14,41)(12,42)
     \put(12,36){\makebox(0,0)[l]{$\scriptstyle p^{}_{\scriptscriptstyle 2}$}}
   \end{picture}
   &
   \begin{picture}(60,90)(0,-15)
     \put(0,0){\framebox(60,60){}}
     
     \put(15,45){\circle{10}}
     \put(15,45){\makebox(0,0){$\scriptstyle l_0$}}
     \put(15,15){\circle{10}}
     \put(15,15){\makebox(0,0){$\scriptstyle l_1$}}
     \put(45,15){\circle{10}}
     \put(45,15){\makebox(0,0){$\scriptstyle l_2$}}
     
     \qbezier(11.5,18.5)(9,22)(9,27)
     \put(9,27){\vector(0,1){6}}
     \qbezier(9,33)(9,38)(11.5,41.5)
     \put(7,30){\makebox(0,6)[r]{$e$}}
     
     \qbezier(18.5,41.5)(21,38)(21,33)
     \put(21,33){\vector(0,-1){6}}
     \qbezier(21,27)(21,22)(18.5,18.5)
     \put(23,30){\makebox(0,6)[l]{$s$}}
     
     \qbezier(18.5,18.5)(22,21)(27,21)
     \put(27,21){\vector(1,0){6}}
     \qbezier(33,21)(38,21)(41.5,18.5)
     \put(29,23){\makebox(6,0)[b]{$s$}}
     
     \qbezier(41.5,11.5)(38,9)(33,9)
     \put(33,9){\vector(-1,0){6}}
     \qbezier(27,9)(22,9)(18.5,11.5)
     \put(29,7){\makebox(6,0)[t]{$e$}}
     
     \put(30,50){\framebox(10,10){$s$}}
     \put(50,30){\framebox(10,10){$e$}}
   \end{picture}
   &
   \begin{picture}(160,90)(0,5)
     \multiput(0,30)(60,0){3}{
       \put(17,40){\synchron}
       \put(20,46){\line(0,1){9}}
       
       \put(17,-6){\synchron}
       \put(20,-6){\line(0,-1){9}}
     }
     \put(0,30){\framebox(40,40){$T_1$}}
     \put(15,60){\framebox(10,10){$b$}}
     \put(30,45){\framebox(10,10){$r$}}
     \put(15,30){\framebox(10,10){$f$}}
     \put(0,45){\framebox(10,10){$p$}}
     
     \put(60,30){\framebox(40,40){$P$}}
     \put(75,60){\framebox(10,10){$s$}}
     \put(75,30){\framebox(10,10){$e$}}
     
     \put(120,30){\framebox(40,40){$T_2$}}
     \put(135,60){\framebox(10,10){$p$}}
     \put(150,45){\framebox(10,10){$b$}}
     \put(135,30){\framebox(10,10){$r$}}
     \put(120,45){\framebox(10,10){$f$}}
     
     \put(20,85){\line(1,0){60}}
     \put(47,85){\trigger}
     \put(50,91){\line(0,1){4}}
     \put(137,85){\anytype}
     \put(140,91){\line(0,1){4}}
     \put(50,95){\line(1,0){90}}
     
     \put(20,15){\line(1,0){60}}
     \put(47,15){\triggerd}
     \put(50,9){\line(0,-1){4}}
     \put(137,9){\anytype}
     \put(140,9){\line(0,-1){4}}
     \put(50,5){\line(1,0){90}}
   \end{picture}
   \\
   $(a)$ & $(b)$ & $(c)$
 \end{tabular}
 \mycaption{BIP models of a task $(a)$ and a processor $(b)$; connectors
   $(c)$}
 \label{fig:task} \label{fig:cpu} \label{fig:system}
\end{figure}

The BIP model of the task component type is shown in \fig[(a)]{task}.  It
has an ``idle'' state $s$, and, for each processor $i \in [1,M]$, a
``compute'' state $c_i$ and a ``wait'' state $w_i$.  An idle task (in state
$s$) can begin execution on the processor $i$ by taking the transition
labeled $b_i$ from the state $s$ to the state $c_i$.  It can finish
execution by taking the transition labeled $f_i$ from the state $c_i$ back
to the state $s$.

A task running on the processor $i$ can be preempted (transition labeled
$p_i$ from the state $c_i$ to the state $w_i$) and, subsequently, resumed
(transition labeled $r_i$ from the state $w_i$ to the state $c_i$).

The BIP model of the processor component type is shown in \fig[(b)]{cpu}.
A processor $i$ is free in the control state $l_0$, and can start executing
a new task by taking a transition labeled $s$ to the state $l_1$.  To do
so, it must synchronize with the ``begin'' port $b_i$ of the task to be
allocated.

From the state $l_1$, the processor can move back to state $l_0$, if the
running task finishes (transition labeled by $e$).  Otherwise, it can
preempt the running task and start a newly arriving task by taking a
transition to $l_2$, labeled by the port $s$.  To do so, it must
synchronize with the ``begin'' port $b_i$ of the newly arrived task and
``preempt'' port $p_i$ of the currently running task.  Similarly, for a
processor with two tasks (state $l_2$) an interaction $e f_i r_i$ ends the
running task and resumes the preempted one.

Each task is connected with every processor and all other tasks.
\fig[(c)]{system} shows the corresponding connectors $[bs]'p$ and $[fe]'r$
between a task $T_1$, a processor $P$, and another task $T_2$.  For the
sake clarity, we show only the relevant ports.

Thus, in a system of $N$ tasks and $M$ processors, there are $2N(N-1)M$
connectors.  We say that connectors are \emph{dense} in this system, which
favors the boolean engine: execution time of the boolean engine is linear
in the number of components (\cf \secn{simulation} and \fig[(b)]{bench}),
whereas that of the enumerative engine is linear in the number of
connectors and quadratic in the number $N$ of tasks.

Observe that \eg connector $[bs]'p$ has two interactions $bs$ and $bsp$.
Whenever a task is already running on a processor, it has to be preempted
before a new one can be started.  This is realized by the maximal progress
rule, \ie giving priority to $bsp$ over $bs$.  Both enumerative and boolean
engines automatically pick the maximal interaction, which does not increase
computational complexity of the underlying algorithms (contrary to
arbitrary priorities).


\subsection{Simulation results} 
\label{sec:simulation}

We measured the engine execution times for both examples for $10^6$
iterations of the engine loop.  \fig[(a)]{bench} shows the engine execution
times obtained with both the enumerative and boolean engines, related to
the number of components in the system.  \fig[(b)]{bench} shows the size of
the system BDD $f_S$.  Observe that for both examples, the size of this BDD
is linear in the number of components.

\begin{figure}
  \centering

  \begin{tabular}{@{}cc@{}}
    \includegraphics[angle=0,width=0.48\textwidth]{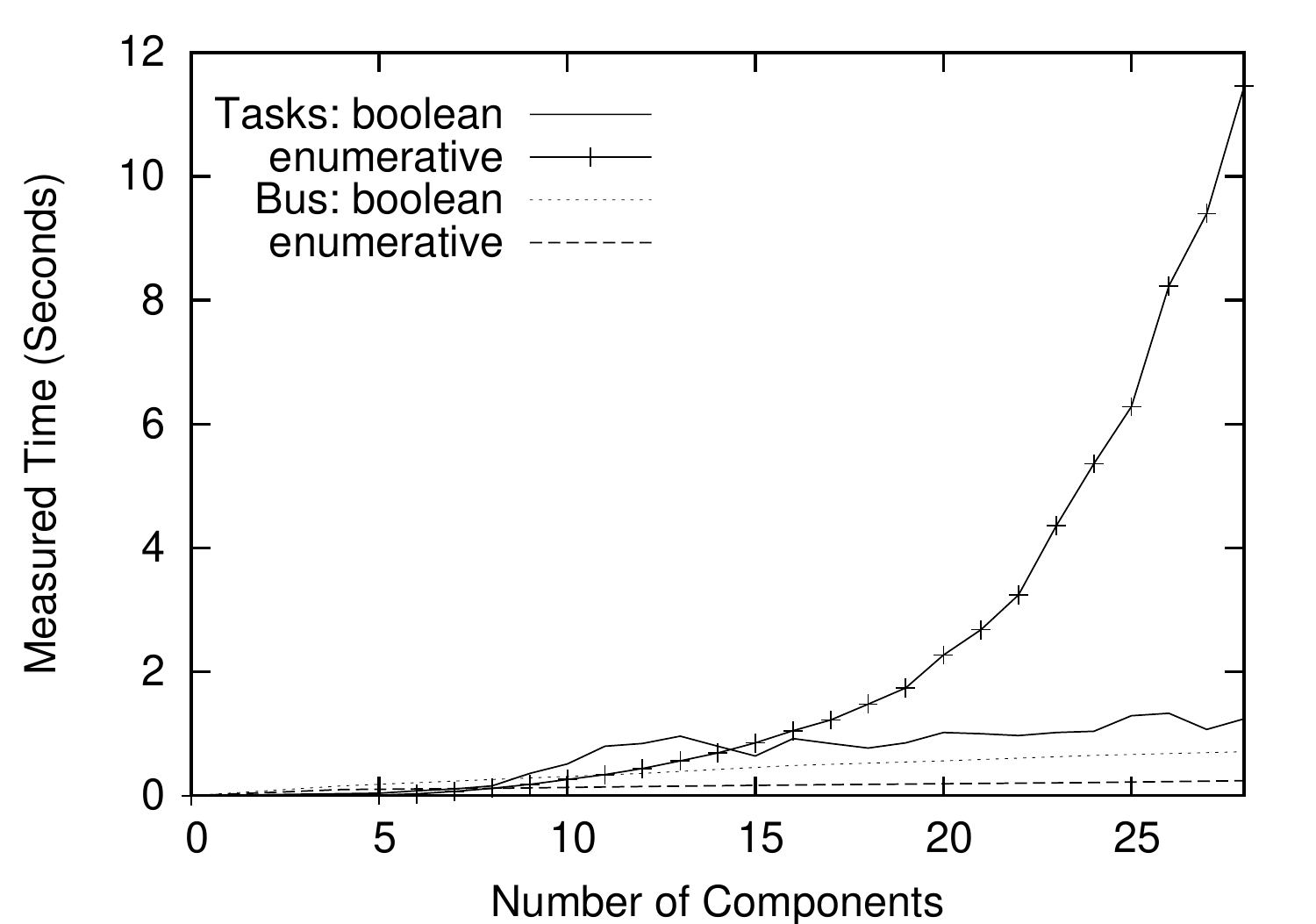}
    &   
    \includegraphics[angle=0,width=0.5\textwidth]{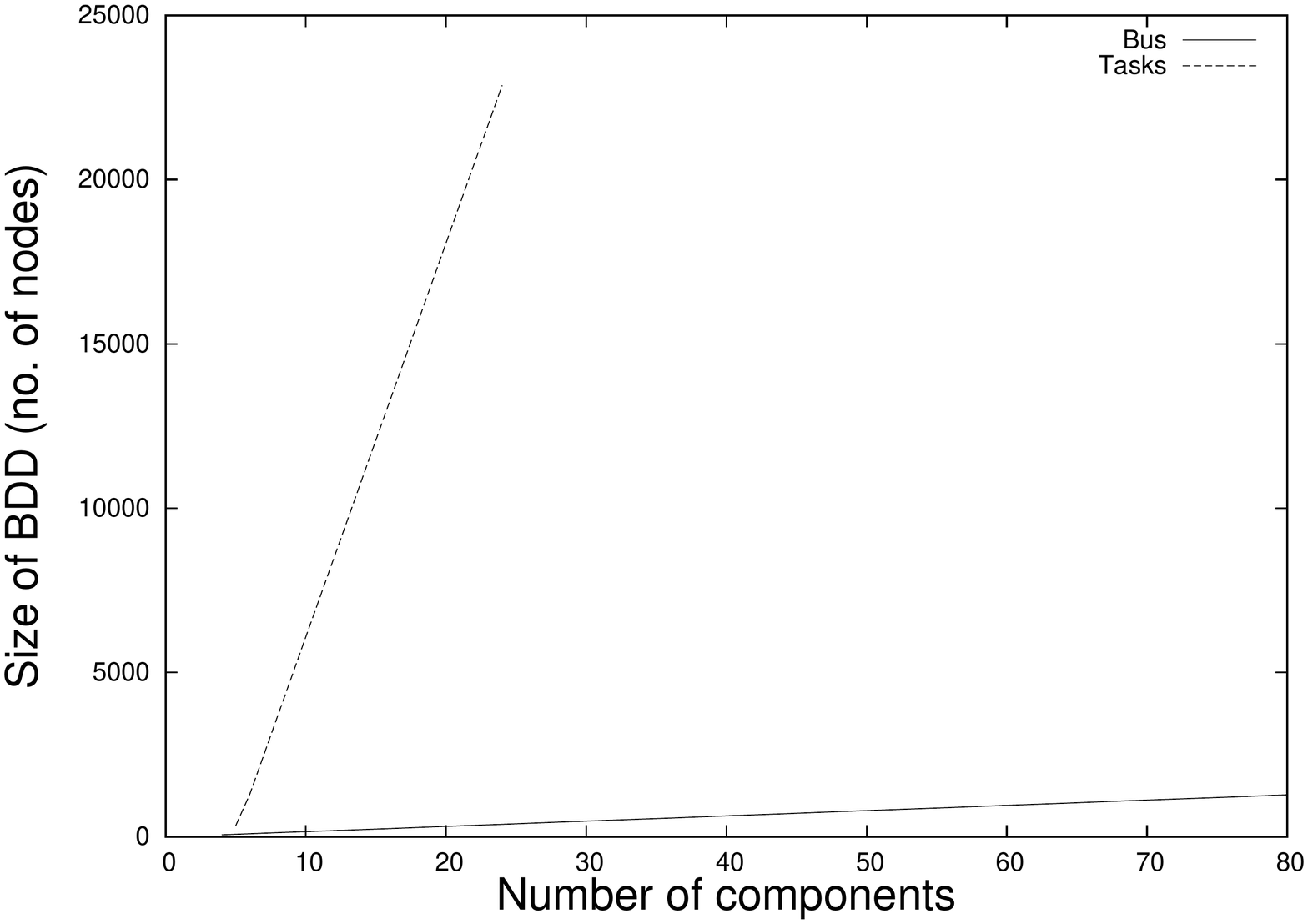}    
    \\
    $(a)$ & $(b)$
  \end{tabular}
  \mycaption{Engine execution times $(a)$ and BDD size $(b)$}
  \label{fig:bench}
\end{figure}

As expected, for the Bus example, the execution times of both engines are
close and linear in the number of components (dashed lines in
\fig[(a)]{bench}).  The enumerative engine outperforms the boolean one.
This is due to the fact that the basic operation of the boolean engine (BDD
conjunction in \eq{protocol}) is more expensive than that of the
enumerative engine (connector evaluation).

In the Preemptable Tasks example, we fixed the number of processors to
$M=4$.  The execution time of the enumerative engine is linear in the
number of connectors, \ie quadratic in the number of components (solid
lines in \fig{bench}).  The execution time of the boolean engine is linear
in the number of components.  Thus boolean engine considerably outperforms
the enumerative one.


\section{Conclusion}

We presented the symbolic implementation of the BIP execution framework.
This implementation is based on computing boolean representation for
components and connectors by using an existing BDD package.  The boolean
representation is used by the engine at runtime to compute the interaction
to be executed at each iteration of the engine loop.  The aim of the
symbolic implementation is to reduce the overhead observed in the original
enumerative engine due to this computation.  The main goal of this paper is
to demonstrate the feasibility of this approach.  Therefore, even though we
provide an implementation technique for priorities, the main focus of this
paper is on the boolean representation of connectors.

We have compared the execution times of the two engines.  For the
enumerative engine, the worst-case complexity of the engine protocol is
proportional to the number of connectors, whereas, for the symbolic
implementation it is proportional to the size of the BDD for the function
$f_S$ representing the system.

The engine execution times were evaluated for two examples favoring
respectively the two engines.  For systems with dense connectors (as in the
Preemptable Tasks example), the execution time of the enumerative engine
explodes, whereas that of the boolean engine remains small due to the small
size of the BDD for $f_S$.  For systems where connectors are sparse (as in
the Bus example), the execution times of both engines are close, with the
enumerative one potentially outperforming the symbolic one.

The size of the BDD is influenced by the order of variables.  Hence, we
alternate port and state variables (\cf \fig[(a)]{bdds}), as these are
strongly correlated\mdash the active ports of each atomic component are
determined by its current state.  Using this variable ordering we obtained
system BDDs linear in the number of components for both examples that we
have considered.

One of the directions for future work is to determine the optimal order of
atomic components depending on the topology of the connectors.  Indeed,
graph-theoretical methods like clique detection could allow keeping
strongly interconnected atomic components close to each other in order to
further reduce the size of the system BDD.

We believe that the techniques presented in this paper can improve the
efficiency of the BIP engine in many practical situations.  Furthermore,
they can be adapted for other frameworks with structured multi-way
communication, like Reo \cite{Arbab05}, Lotos \cite{lotos}, etc.


\section*{Acknowledgements}
The authors would like to thank Marius Bozga, Joseph Sifakis, and
Chaouki Zerrari for constructive remarks and valuable discussion
concerning symbolic implementation of priority models.  Marius' advice
on the usage of BDDs was crucial to the completion of this work.

\bibliographystyle{eptcs}
\bibliography{bip,connectors,causal,latex8}

\end{document}

%% file: system-bdd.pdf_t
\begin{picture}(0,0)%
\includegraphics{system-bdd.pdf}%
\end{picture}%
\setlength{\unitlength}{2486sp}%
\begingroup\makeatletter\ifx\SetFigFontNFSS\undefined%
\gdef\SetFigFontNFSS#1#2#3#4#5{%
  \reset@font\fontsize{#1}{#2pt}%
  \fontfamily{#3}\fontseries{#4}\fontshape{#5}%
  \selectfont}%
\fi\endgroup%
\begin{picture}(2052,3305)(-14,-2998)
\put(  1,-961){\makebox(0,0)[lb]{\smash{{\SetFigFontNFSS{7}{8.4}{\rmdefault}{\mddefault}{\updefault}{\color[rgb]{0,0,0}$P_2$}%
}}}}
\put(  1,-16){\makebox(0,0)[lb]{\smash{{\SetFigFontNFSS{7}{8.4}{\rmdefault}{\mddefault}{\updefault}{\color[rgb]{0,0,0}$P_1$}%
}}}}
\put(  1,-16){\makebox(0,0)[lb]{\smash{{\SetFigFontNFSS{7}{8.4}{\rmdefault}{\mddefault}{\updefault}{\color[rgb]{0,0,0}$P_1$}%
}}}}
\put(  1,-16){\makebox(0,0)[lb]{\smash{{\SetFigFontNFSS{7}{8.4}{\rmdefault}{\mddefault}{\updefault}{\color[rgb]{0,0,0}$P_1$}%
}}}}
\put(  1,-1411){\makebox(0,0)[lb]{\smash{{\SetFigFontNFSS{7}{8.4}{\rmdefault}{\mddefault}{\updefault}{\color[rgb]{0,0,0}$Q_2$}%
}}}}
\put(  1,-516){\makebox(0,0)[lb]{\smash{{\SetFigFontNFSS{7}{8.4}{\rmdefault}{\mddefault}{\updefault}{\color[rgb]{0,0,0}$Q_1$}%
}}}}
\put(  1,-2289){\makebox(0,0)[lb]{\smash{{\SetFigFontNFSS{7}{8.4}{\rmdefault}{\mddefault}{\updefault}{\color[rgb]{0,0,0}$Q_n$}%
}}}}
\put(1756,-421){\makebox(0,0)[lb]{\smash{{\SetFigFontNFSS{7}{8.4}{\rmdefault}{\mddefault}{\updefault}{\color[rgb]{0,0,0}$q_1$}%
}}}}
\put(1756,-1321){\makebox(0,0)[lb]{\smash{{\SetFigFontNFSS{7}{8.4}{\rmdefault}{\mddefault}{\updefault}{\color[rgb]{0,0,0}$q_2$}%
}}}}
\put(1756,-2221){\makebox(0,0)[lb]{\smash{{\SetFigFontNFSS{7}{8.4}{\rmdefault}{\mddefault}{\updefault}{\color[rgb]{0,0,0}$q_n$}%
}}}}
\end{picture}%

%% file: states-bdd.pdf_t
\begin{picture}(0,0)%
\includegraphics{states-bdd.pdf}%
\end{picture}%
\setlength{\unitlength}{2486sp}%
\begingroup\makeatletter\ifx\SetFigFontNFSS\undefined%
\gdef\SetFigFontNFSS#1#2#3#4#5{%
  \reset@font\fontsize{#1}{#2pt}%
  \fontfamily{#3}\fontseries{#4}\fontshape{#5}%
  \selectfont}%
\fi\endgroup%
\begin{picture}(1387,2721)(-7,-1875)
\put(856,-196){\makebox(0,0)[b]{\smash{{\SetFigFontNFSS{7}{8.4}{\rmdefault}{\mddefault}{\updefault}{\color[rgb]{0,0,0}$q_2$}%
}}}}
\put(1216,-1096){\makebox(0,0)[b]{\smash{{\SetFigFontNFSS{7}{8.4}{\rmdefault}{\mddefault}{\updefault}{\color[rgb]{0,0,0}$q_n$}%
}}}}
\put(496,659){\makebox(0,0)[b]{\smash{{\SetFigFontNFSS{7}{8.4}{\rmdefault}{\mddefault}{\updefault}{\color[rgb]{0,0,0}$q_1$}%
}}}}
\end{picture}%